%% file: template.tex
\documentclass{Interspeech}

\usepackage{url}
\usepackage{hyperref}

\interspeechcameraready

\title{Unsupervised Rhythm and Voice Conversion to Improve ASR on Dysarthric Speech}

\author[affiliation={1,2 }]{Karl}{El Hajal}
\author[affiliation={1}]{Enno}{Hermann}
\author[affiliation={1}]{Sevada}{Hovsepyan}
\author[affiliation={1}]{Mathew}{Magimai.-Doss}

\affiliation{Idiap Research Institute}{CH-1920 Martigny}{Switzerland}
\affiliation{EPFL, École polytechnique fédérale de Lausanne}{CH-1015 Lausanne}{Switzerland}

\email{\{karl.elhajal,enno.hermann,sevada.hovsepyan,mathew\}@idiap.ch}

\usepackage{comment}

\usepackage{tikz}
\usetikzlibrary{shapes.geometric, arrows}

\usepackage{subcaption}

\usepackage{booktabs}

\begin{document}

\maketitle

\begin{abstract}
    
    Automatic speech recognition (ASR) systems struggle with dysarthric speech due to high inter-speaker variability and slow speaking rates. To address this, we explore dysarthric-to-healthy speech conversion for improved ASR performance. Our approach extends the Rhythm and Voice (RnV) conversion framework by introducing a syllable-based rhythm modeling method suited for dysarthric speech. We assess its impact on ASR by training LF-MMI models and fine-tuning Whisper on converted speech. Experiments on the Torgo corpus reveal that LF-MMI achieves significant word error rate reductions, especially for more severe cases of dysarthria, while fine-tuning Whisper on converted data has minimal effect on its performance. These results highlight the potential of unsupervised rhythm and voice conversion for dysarthric ASR. Code available at: \url{https://github.com/idiap/RnV}.
\end{abstract}

\section{Introduction}

Motor speech impairments like dysarthria can significantly hinder communication by affecting multiple aspects of speech production, including rhythm and articulation~\cite{Duffy2012}. As a result, Automatic Speech Recognition (ASR) systems trained on typical speech often struggle to process dysarthric speech accurately~\cite{Moore2018}. This creates a need for specialized ASR systems that accommodate the unique speech patterns of individuals with dysarthria. However, developing such systems is challenging for two main reasons. First, dysarthric speech deviates significantly from typical speech patterns, and exhibits substantial inter-speaker variability. Speech recognition models particularly struggle with the slower speaking rates common in dysarthria. Second, the scarcity of dysarthric speech data complicates development, as collecting speech samples can be physically demanding for individuals with dysarthria.

Recent studies have explored promising avenues such as the use of synthesis and conversion methods for data augmentation and dysarthric-to-healthy speech conversion~\cite{Soleymanpour2022,Hermann2023,Leung2024}. Among these, unsupervised techniques leveraging Self-Supervised Learning (SSL) speech representations have demonstrated promise due to their zero-shot capabilities and low data requirements. In previous work, we presented an unsupervised Rhythm and Voice (RnV) conversion framework \cite{hajal2025unsupervisedrhythmvoiceconversion} that modifies dysarthric speech to resemble healthy speech, showing promise in improving ASR performance. Indeed, evaluations on the Torgo corpus \cite{Rudzicz2012_torgo} showed that rhythm conversion was particularly beneficial for speakers with more severe dysarthria, improving ASR performance on models trained solely on healthy speech. While RnV enables unsupervised and data-efficient dysarthric speech conversion, the rhythm modeling approach used was not specifically adapted to dysarthric speech, resulting in imprecise segmentation and rhythm modification. Furthermore, while ASR performance improved, it remained unsatisfactory, and the impact of training or adapting ASR models on converted speech was not explored.

\input{figures/rnv-overview}

To address these limitations, this paper explores a syllable-based rhythm modeling approach aimed at improving segmentation and adaptation for dysarthric speech. We investigate the impact of training and adapting ASR models on converted speech by training an LF-MMI~\cite{Povey2016} model and fine-tuning Whisper~\cite{whisper}, providing a comparative analysis of their performance.

\section{Background}

The RnV framework (Fig.~\ref{fig:rnv-overview}) converts dysarthric speech into healthy speech in unsupervised fashion by leveraging properties of self-supervised speech representations \cite{hajal2025unsupervisedrhythmvoiceconversion}. Rhythm conversion is achieved through a modified version of Urhythmic \cite{urhythmic}, replacing soft units with discrete speech representations and extending the any-to-one conversion approach to any-to-any using a general-purpose vocoder. This method first segments speech into three types: silences, sonorants, and obstruents. This is done by clustering discrete units from a speech dataset into 100 centroids, which are then hierarchically grouped. The group with the greatest overlap with silences is labeled as Silence, the one most associated with voiced sections as Sonorants, and the remaining type as Obstruents. Segmentation is performed by comparing each discrete unit to the centroids and computing its log-probability of belonging to each class. A dynamic programming algorithm then merges consecutive units into longer segments, with a parameter $\gamma$ controlling segment length. 
This segmentation enables the calculation of speaking rate and duration distributions for each speech type per speaker. These can be used to modify rhythm through time-stretching, either at the utterance level (global) or at the level of individual segments and speech types (fine-grained).
Voice conversion, on the other hand, employs kNN-VC \cite{baas23_interspeech} to map the phonetic content of the source speaker to the closest matching units of a target speaker.

\section{Methods}

    In this work, we extend the rhythm conversion module of the RnV framework by combining the unsupervised clustering-based method with syllable segmentation and modeling. We further train and adapt ASR models on the converted speech to assess more thoroughly whether conversion helps improve recognition performance.

\subsection{Syllable-based rhythm modeling}

Syllables are fundamental to rhythm modeling in speech. Indeed, analyzing syllable durations helps characterize a speaker’s unique rhythmic patterns \cite{pfau_speaking_rate}. While time-aligned transcriptions are ideal for precise syllable segmentation and speaking rate calculation, alternative methods are needed when such transcriptions are unavailable. Prior work using the Urhythmic method estimates speaking rate by segmenting and counting sonorants per second, as sonorants serve as syllable nuclei \cite{hajal2025unsupervisedrhythmvoiceconversion}.~However, obtaining precise sonorant segments proved challenging due to the unique characteristics of dysarthric speech. To address this, we aim to improve rhythm modeling by directly obtaining syllable segments in an unsupervised manner.

The importance of syllable-level analysis, particularly segmentation and feature extraction, has been demonstrated in recent research on dysarthric speech detection \cite{hovsepyan_syllabel_features}. Indeed, such approaches mirror human auditory processing, where sound is initially decomposed into frequencies by the cochlea and subsequently segmented into syllables by the cortex. Inspired by this work, we adopt a segmentation approach which leverages theta oscillations and the sonority envelope \cite{RASANEN2018130}.  This method segments syllables based on either syllable onsets (valleys in the envelope) or syllable nuclei (typically vowels, represented by peaks in the envelope). By extracting peak and valley timestamps, we can define syllable segments as either valley-to-valley or peak-to-peak intervals.

While effective, this method struggles with noisy speech, as noise can introduce false peaks and valleys. Given the presence of noise in the data which we will be using for evaluation, we incorporate a filtering step using the initial discrete unit clustering method. This allows us to ignore segments that fall outside speech regions. Specifically, we define speech segments as the combined obstruent and sonorant regions, while silence segments are classified as non-speech. This segmentation acts as a form of Voice Activity Detection (VAD), helping to distinguish speech from noise. In preliminary experiments, we observed that this approach was more robust than traditional VAD methods in identifying non-speech regions in the context of dysarthric speech. Any peak or valley within a non-speech region is discarded, improving segmentation accuracy. The segmentation steps are visualized in Figure \ref{fig:segmentation_steps}.

After filtering, we focus on syllable nuclei (peaks), as they are less prone to false positives. By measuring peak-to-peak syllable durations for each speaker, we establish two rhythm modeling and conversion approaches:
\\
\textbf{Global}: We calculate the syllables-per-second rate for each speaker to determine a global speaking rate. To convert a source speaker’s utterance to a target speaker’s rhythm, we time-stretch the discrete units at the utterance level using the ratio of their speaking rates. \input{figures/segmentation_steps}
\\ 
\textbf{Fine-Grained}:  Inspired by the fine-grained Urhythmic approach, we fit syllable durations to a gamma distribution, creating a speaker-specific duration model. To convert rhythm, we map each source syllable segment’s duration to the target distribution using the Cumulative Distribution Function (CDF) and the Percent Point Function (PPF). This ensures that each segment's duration maintains its probability rank within the target distribution, preserving natural rhythm characteristics.

\subsection{ASR adaptation}

In this study, we investigate whether speech conversion enhances ASR performance by training and adapting models on the converted data. To achieve this, we explore two approaches: first, we train LF-MMI ASR models from scratch, and second, we fine-tune a pre-trained Whisper model originally trained on healthy speech. A detailed explanation of these methodologies is provided in the following section.

\section{Experimental Setup}

\subsection{RnV implementation}

We implement the framework similarly to \cite{hajal2025unsupervisedrhythmvoiceconversion}. We use the 6th layer of WavLM Large \cite{wavlm} as our speech representation, and reconstruct waveforms using a pre-trained HiFi-GAN V1 vocoder \cite{hifigan} checkpoint trained using the pre-matched paradigm from \cite{baas23_interspeech}. 
For the clustering-based segmentation, we use $\gamma = 3$. For kNN-VC, we find the $k=8$ nearest units calculated using the cosine distance, and apply weighted averaging to the obtained units.

\subsection{Datasets}
Our evaluation is conducted using the Torgo corpus \cite{Rudzicz2012_torgo}, which contains voice samples from 15 participants: 8 individuals with dysarthria (stemming from either Cerebral Palsy or Amyotrophic Lateral Sclerosis) and 7 control subjects. Speakers with dysarthria are classified into four severity levels: severe, moderately severe, moderate, and mild. The collected samples encompass 725 sentences and 2,340 isolated words. We use both the recordings from the head-mounted and array microphones. For the target speech in our conversion process, we use the LJSpeech database~\cite{ljspeech}, which features 24 hours of single-speaker English audiobook narration. We process all audio samples by standardizing them to 16kHz, which is WavLM's expected input sampling rate, and normalizing volume levels to -20dB.

\input{figures/speaking_rates_plot}

\subsection{ASR experiments}

We use a Leave-One-Speaker-Out approach, where to evaluate the performance on each speaker, we train/fine-tune each model on the data from all other speakers and test on the remaining speaker.
To evaluate the impact of different conversion setups on ASR performance, we conduct experiments with two different, existing models. The proposed conversion methods could also be combined with other ASR models for potential better absolute performance, but we emphasized openly available implementations in this work.

\textbf{LF-MMI}: We train factorized time-delay neural network \cite{Povey2018} acoustic models with the sequence-discriminative LF-MMI loss~\cite{Povey2016}. The models are trained in Kaldi~\cite{Povey2011} using the training recipe from~\cite{enno_lfmmi}, i.e. first training HMM-GMM ASR models and then using their alignments for LF-MMI training with speed perturbation (factors 0.9, 1.0, 1.1). The only difference is that we do not use i-vectors for simplicity. Also following~\cite{enno_lfmmi}, when decoding isolated words a grammar restricts the output to one of the possible options, for sentences a bigram language model is trained on all sentence data.

\textbf{Whisper}: We fine-tune the pre-trained Whisper base model \cite{whisper}, which comprises 74M parameters. It is pre-trained on a diverse multilingual dataset for generalization across languages, and incorporates a multitask learning framework which includes transcription, translation, and language identification. It employs a sequence-to-sequence approach, where audio inputs are converted into log-Mel spectrograms and processed by a convolutional encoder to extract features. These features are then passed through a transformer-based encoder, which captures long-range dependencies. A transformed decoder generates text tokens autoregressively, conditioned on the encoded audio representations. For fine-tuning, we use a batch size of 32, a learning rate $\alpha=1e-5$, and early stopping with patience set to 5 epochs. There is no additional language model.

ASR performance is assessed using the word error rate (WER). First, we present the speaker-averaged WER results for Torgo speakers with dysarthria across all conversion setups. We then plot per-speaker WER results for selected setups for a more detailed analysis. Following the approach in \cite{enno_lfmmi}, we report the results separately for isolated words and sentences, as each scenario presents distinct challenges.

\input{figures/syllable_distributions_plot}
\subsection{Conversion setups}

For conversion, similarly to \cite{hajal2025unsupervisedrhythmvoiceconversion}, we first train the Urhythmic segmenter on LJSpeech to obtain the 100 centroids, hierarchically group them into the three speech types, and perform segmentation as described in Section 2. Using both the Urhythmic and syllable-based methods, we compute global and fine-grained rhythm models for LJSpeech and each Torgo speaker, enabling the conversion of Torgo utterances to LJSpeech under different setups. We then evaluate and compare ASR performance across original, vocoded (encoded and decoded without modification), voice-converted, rhythm-converted (using each global and fine-grained method), and rhythm + voice-converted samples. In the next section, we refer to the rhythm conversion approaches as Syllable and Urhythmic, denoting the fine-grained and global methods as Fine and Global, respectively.

\input{tables/dysarthric_wer}
\input{figures/per_speaker_wer_plots}

\section{Results}

Figure \ref{fig:speaking_rates} presents the speaking rates calculated for each Torgo speaker using the syllable-based method. We can observe that speaking rates increase with lower severity levels as expected. Severe and moderately severe speakers exhibit rates around 2 syllables per second, while control speakers have a rate close to 4 syllables per second, which aligns with the typical average. Compared to speaking rates derived from counting sonorants per second, as reported in \cite{hajal2025unsupervisedrhythmvoiceconversion}, the syllable-based method provides clearer separation between severity levels and more consistent rates for control speakers.
Moreover, Figure \ref{fig:syllable_distributions} illustrates the syllable duration gamma distributions for control speaker MC02 and speaker with dysarthria M02. The probability density for the control speaker peaks just below 0.25 seconds, which would correspond to a rate of 4 syllables per second. In contrast, the distribution for the speaker with dysarthria is more variable, peaks at 0.4 seconds, and has a longer tail, with some syllable durations exceeding 1 second.

Table \ref{tab:wer_results} presents the WER results averaged across dysarthric Torgo speakers for both the LF-MMI and fine-tuned Whisper-base models. For the Whisper-base models, the conversion methods do not yield improvements over fine-tuning on the original data, with WERs of approximately 30 for sentences and 50 for isolated words. In contrast, the LF-MMI model shows clear improvement. When trained on the original data, its performance is comparable to the Whisper models. Applying kNN-VC significantly enhances results, achieving the lowest WER for isolated words and reducing the WER to 18.4 for sentences. Rhythm conversion methods also improve performance: both global rhythm conversion approaches reduce WER to around 20 for sentences, while the syllable-based fine-grained method outperforms the Urhythmic-based fine-grained method and performs similarly to the global methods. Finally, combining rhythm conversion with kNN-VC further reduces WER, achieving the best overall sentence performance (15.9) when either global rhythm conversion method is used. For isolated words, this combination performs comparably to kNN-VC alone and outperforms rhythm conversion alone.

Figure \ref{fig:wer_plots} presents the WER results per speaker for different configurations. For sentences, the vocoded data interestingly outperforms the original data. Both kNN-VC and global syllable-based conversion show clear improvements over the original and vocoded data, particularly for speakers with more severe dysarthria. Combining these methods further reduces WER, except for speaker M04, where rhythm conversion has minimal impact. Performance for mild and control speakers on the other hand is largely unaffected or slightly improved by the conversion methods. For isolated words, kNN-VC provides the most significant improvements, while its combination with rhythm modeling yields additional gains in some cases, though the effect is less significant than for sentences.

\section{Discussion and conclusions}

The rhythm analysis and modeling results demonstrate that syllable-based segmentation is well-suited for dysarthric speech. The clear correlation between speaking rate and dysarthria severity supports this approach, as speaking rate increases with lower severity. Additionally, fitting a gamma distribution to each speaker's syllable durations provides a more detailed representation of individual rhythm characteristics. Beyond conversion, the ability to capture detailed rhythmic information could be helpful for diagnostic and assessment tools for dysarthric speech.

Dysarthric-to-healthy rhythm conversion proved particularly beneficial for the LF-MMI model, leading to noticeable WER reductions, especially for speakers with severe and moderately severe dysarthria. Global rhythm modeling methods performed best, as they rely on simple utterance-level time-stretching, minimizing errors and preventing artifacts. Further, these methods do not require highly precise speaking rate calculations, as long as the overall speaking rate is increased for highly severe cases. Among fine-grained rhythm modeling methods, the syllable-based approach was more effective than the fine-grained Urhythmic method. While the former achieved performance improvements comparable to global methods, the latter introduced artifacts due to imprecise segmentation, resulting in a drop in ASR performance. In addition, we observed that voice conversion using kNN-VC was just as effective in improving the LF-MMI model's performance. This is likely due to reducing the inter-speaker variations, which simplifies the task for the model. The best results were obtained by performing both rhythm and voice conversion, demonstrating that both techniques are complementary.

On the other hand, fine-tuned Whisper-base models did not benefit from rhythm or voice conversion, likely due to their extensive pre-training on over 680,000 hours of healthy speaker audio.~This large-scale training enables Whisper to generalize well across speakers, making voice standardization unnecessary. Furthermore, once fine-tuned on dysarthric speech, the model's transformer architecture can adapt to slower speaking rates, reducing the need for rhythm conversion.~Indeed, the fine-tuned models significantly outperform the pre-trained base model, which, as reported in \cite{hajal2025unsupervisedrhythmvoiceconversion}, produced hallucinated outputs when not adapted to dysarthric speech. These hallucinations are caused by slower speaking rates and resulted in excessively high WER values for speakers with severe dysarthria.
In contrast, the LF-MMI model benefited from rhythm and voice conversion as it was trained from scratch. Converting all Torgo data to a single speaker reduced inter-speaker variations, simplifying the training process and enhancing test-time performance on similarly converted data.
Future work could focus on identifying distinct syllable groups to enable more detailed rhythm modeling.

\section{Acknowledgements}
This work was partially supported by the Swiss National Science Foundation (SNSF) through the project ``Pathological Speech Synthesis (PaSS)'' (grant agreement no. 219726), by the SNSF through the Bridge Discovery project ``Emotion in the loop - a step towards a comprehensive closed-loop deep brain stimulation in Parkinson’s disease (EMIL)'' (grant agreement no. 40B2–0\_194794), and by the Innosuisse through the flagship project ``Inclusive Information and Communication Technologies (IICT)'' (grant agreement no. PFFS-21-47).

\bibliographystyle{IEEEtran}
\bibliography{mybib}

\end{document}

%% file: figures/rnv-overview.tex
\begin{figure}[]
    \centering
    
    \vspace{-0.5em}
    \resizebox{\columnwidth}{!}{%
    \begin{tikzpicture}[node distance=0cm]

    \tikzset{
    block/.style={rectangle, rounded corners, minimum width=1.5cm, minimum height=1cm, text centered, draw=black, fill=gray!05, font=\bfseries},
    arrow/.style={thick,->,>=stealth},
    data/.style={rectangle, minimum width=1.5cm, minimum height=0.8cm, text centered, draw=black, fill=blue!20, font=\bfseries}
}
    
        \node (input) [data, align=center] {Input\\ Speech};
        \node (ssl) [block, right of=input, xshift=2cm, align=center] {SSL\\ Encoder};
        \node (rhythm) [block, right of=ssl, xshift=2cm, align=center] {Rhythm\\ Converter};
        \node (knnvc) [block, right of=rhythm, xshift=2cm, align=center] {kNN-VC};
        \node (vocoder) [block, right of=knnvc, xshift=2cm, align=center] {Vocoder};
        \node (output) [data, right of=vocoder, xshift=2cm, align=center] {Converted\\ Speech};
    
        \draw [arrow] (input) -- (ssl);
        \draw [arrow] (ssl) -- (rhythm);
        \draw [arrow] (rhythm) -- (knnvc);
        \draw [arrow] (knnvc) -- (vocoder);
        \draw [arrow] (vocoder) -- (output);
    
        \node (sourceRhythm) [data, below of=rhythm, yshift=-1.5cm, align=center] {Source\\Rhythm\\Model};
        \node (targetRhythm) [data, above of=rhythm, yshift=1.5cm, align=center] {Target\\Rhythm\\Model};
        \node (targetDB) [data, above of=knnvc, yshift=1.5cm, align=center] {Target\\Unit\\Database};
    
        \draw [arrow] (sourceRhythm) -- (rhythm);
        \draw [arrow] (targetRhythm) -- (rhythm);
        \draw [arrow] (targetDB) -- (knnvc);
    
    \end{tikzpicture}
    }
    \caption{RnV Framework Overview}
    \label{fig:rnv-overview}
\end{figure}
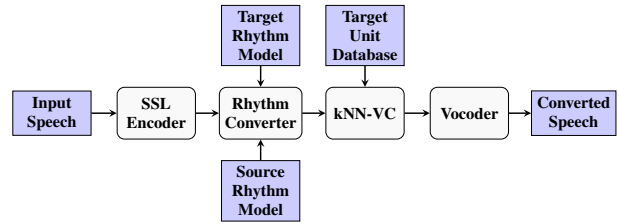

%% file: figures/segmentation_steps.tex
\begin{figure}[h]
    \centering
    \includegraphics[width = \columnwidth]{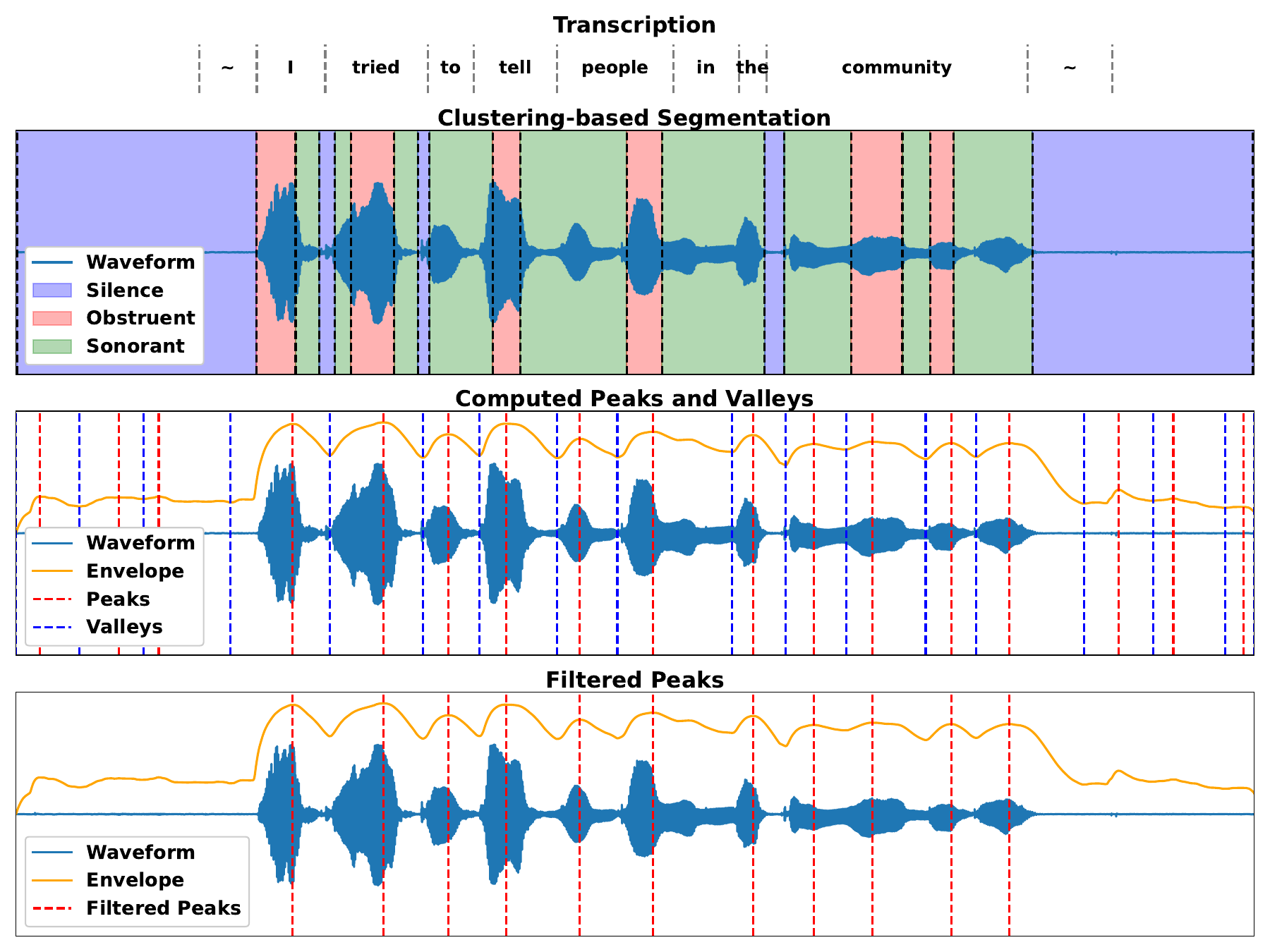}
    \caption{Segmentation steps for Torgo speaker M02 pronouncing 'I tried to tell people in the community'.}
    \label{fig:segmentation_steps}
\end{figure}

%% file: figures/speaking_rates_plot.tex
\begin{figure}[h]
    \centering
    \includegraphics[width = .7\columnwidth]{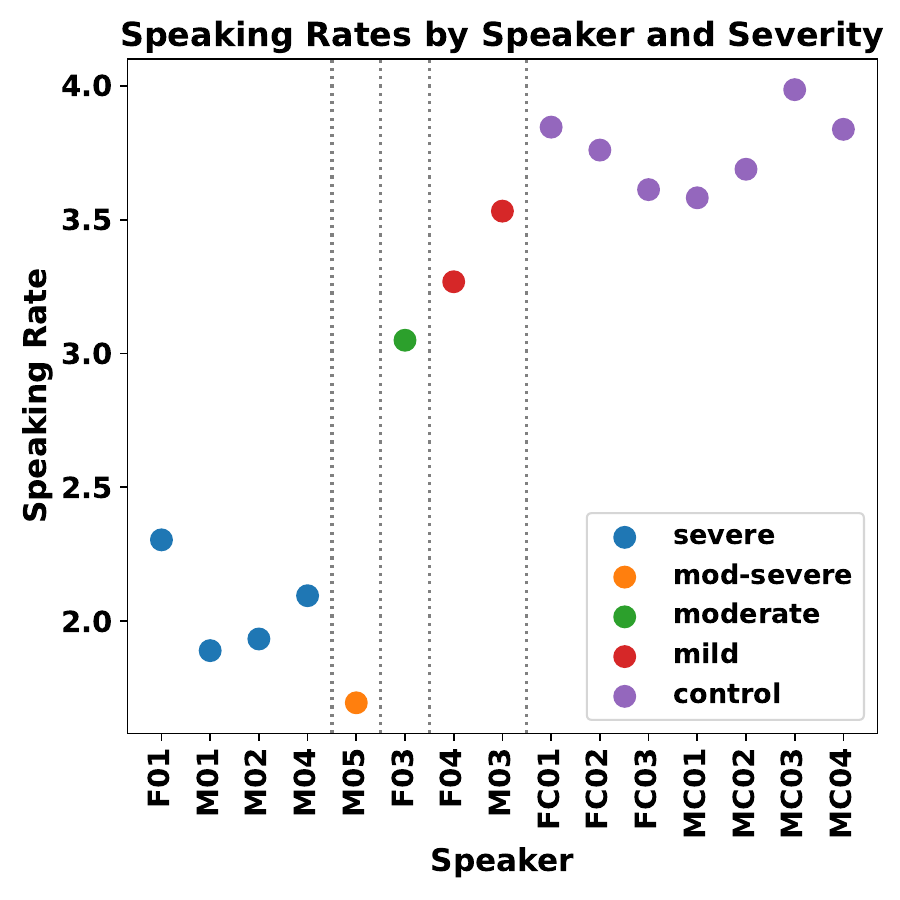}
    \caption{Global speaking rates computed using the Syllable-based method for each Torgo speaker, categorized by severity.}
    \label{fig:speaking_rates}
\end{figure}

%% file: figures/syllable_distributions_plot.tex
\begin{figure}[h]
    \centering
    \includegraphics[width = .7\columnwidth]{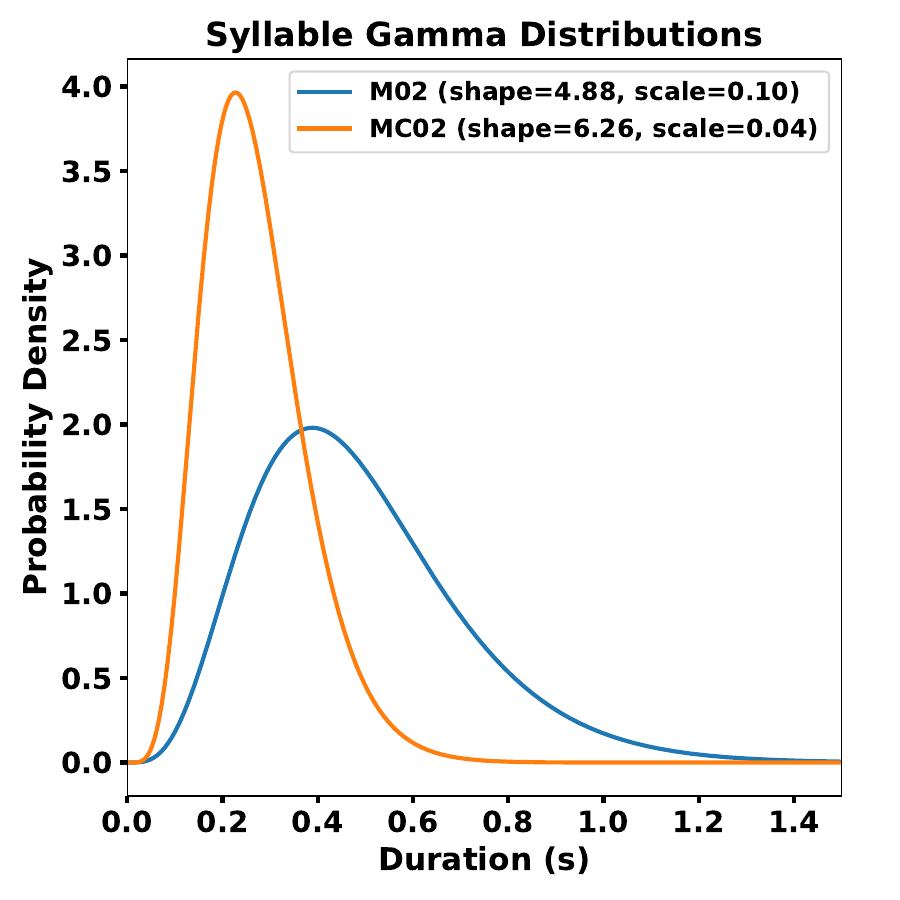}
    \caption{Comparison of syllable gamma duration distributions for control speaker MC02 and dysarthric speaker M02.}
    \label{fig:syllable_distributions}
\end{figure}

%% file: tables/dysarthric_wer.tex
\begin{table}
\centering
\caption{WER Results averaged over Dysarthric Torgo speakers for all conversion setups}
\label{tab:wer_results}
\small
\resizebox{\columnwidth}{!}{
\begin{tabular}{l cc | cc}
\toprule
 & \multicolumn{2}{c|}{\textbf{LF-MMI}} & \multicolumn{2}{c}{\textbf{Fine-tuned Whisper}} \\
\textbf{Experiment} & \textbf{Isolated} & \textbf{Sentences} & \textbf{Isolated} & \textbf{Sentences} \\
\midrule
Original                        & 44.8 & 31.2 & 48.35 & 29.62 \\
Vocoded                         & 43.3 & 24.8 & 50.66 & 32.38 \\
kNN-VC                          & \textbf{37.6} & 18.4 & 53.18 & 34.53 \\
Urhythmic (Fine)                & 60.0 & 26.2 & 54.67 & 37.21 \\
Urhythmic (Global)              & 42.6 & 20.6 & 49.51 & 30.09 \\
Syllable (Fine)                 & 46.5 & 20.6 & 60.01 & 33.94 \\
Syllable (Global)               & 43.8 & 19.4 & 50.93 & 31.81 \\
Urhythmic (Fine) + kNN-VC       & 52.1 & 17.9 & 56.26 & 34.10 \\
Urhythmic (Global) + kNN-VC     & 38.4 & \textbf{15.9} & 51.19 & 39.49 \\
Syllable (Fine) + kNN-VC        & 39.4 & 16.9 & 57.73 & 33.32 \\
Syllable (Global) + kNN-VC      & 39.0 & \textbf{15.9} & 52.75 & 36.29 \\
\bottomrule
\end{tabular}
}
\end{table}

%% file: figures/per_speaker_wer_plots.tex
\begin{figure*}[t]
  \centering
  \begin{subfigure}{1.01\columnwidth}
    \includegraphics[width=\columnwidth]{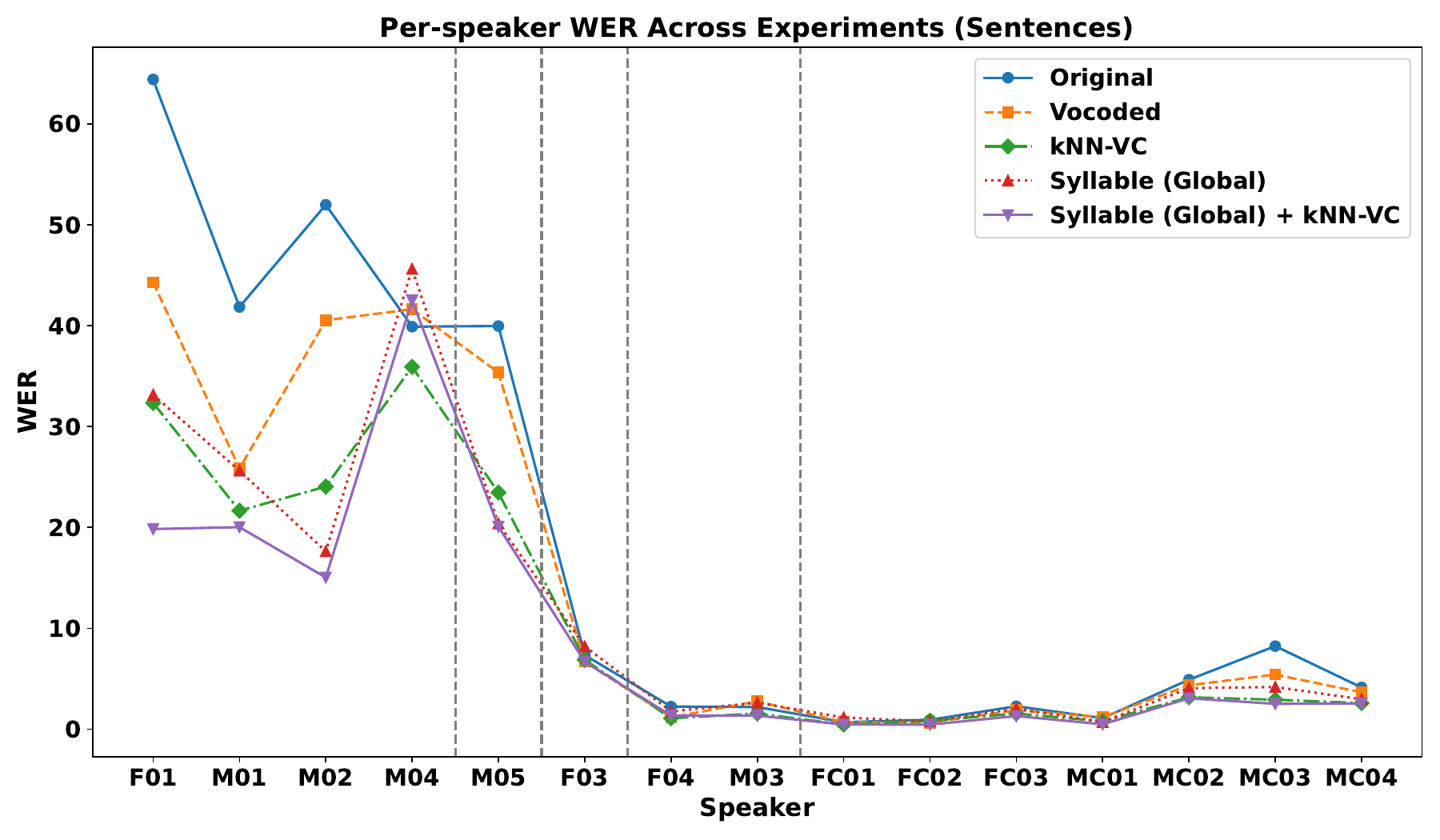}
    \caption{Sentences}
    \label{fig:sentences_wer}
  \end{subfigure}
  \begin{subfigure}{1.01\columnwidth}
    \includegraphics[width=\columnwidth]{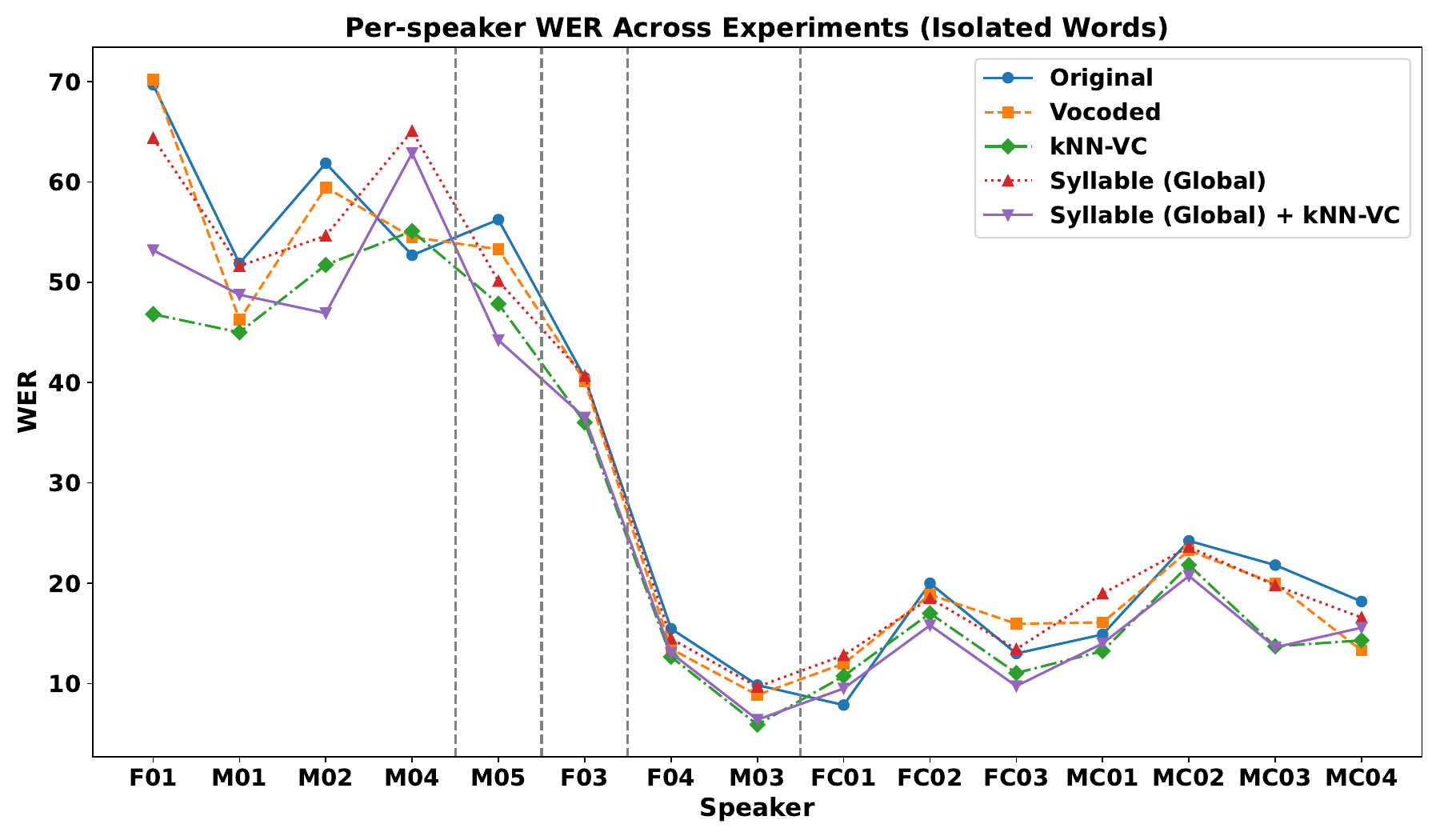}
    \caption{Isolated Words}
    \label{fig:isolated_words_wer}
  \end{subfigure}
  
  \caption{Per-speaker WER results on Torgo using the LF-MMI model for different conversion setups.}
  \label{fig:wer_plots}
\end{figure*}